\documentclass[aps,prl,twocolumn,showpacs,preprintnumbers,superscriptaddress,dblfloatfix]{revtex4-1}
\usepackage{graphicx}   
\usepackage{dcolumn}   
\usepackage{bm}        
\usepackage{amssymb}    
\usepackage{setspace}
\usepackage{amsmath, amssymb, setspace}
\usepackage{array}
\usepackage{booktabs}
\usepackage{mathrsfs}
\usepackage{indentfirst}
\usepackage{slashed}
\usepackage{float}
\usepackage{lmodern}
\usepackage{hyperref}
\usepackage{xcolor}
\usepackage{multirow}
\usepackage{epstopdf}
\usepackage{ulem}
\usepackage{tabularx,lipsum}
\usepackage[justification=raggedright]{caption}
\usepackage[flushleft]{threeparttable}
\usepackage{hyperref}
\usepackage{soul}
\usepackage{subfig}

\begin{document}

\title{Positrons and 511 keV radiation as tracers of recent binary neutron star mergers}

\preprint{IPMU18-0173}

\author{George M. Fuller}
\email[]{gfuller@ucsd.edu}
\affiliation{Department of Physics, University of California, San Diego\\
La Jolla, California 92093-0354, USA}

\author{Alexander Kusenko}
\email[]{kusenko@ucla.edu}
\affiliation{Department of Physics and Astronomy, University of California, Los Angeles\\
Los Angeles, CA 90095-1547, USA}
\affiliation{Kavli Institute for the Physics and Mathematics of the Universe (WPI), UTIAS\\
The University of Tokyo, Kashiwa, Chiba 277-8583, Japan}

\author{David Radice}
\email[]{dradice@ias.edu}
\affiliation{
Institute for Advanced Study, Princeton, NJ 08540, USA}
\affiliation{Department of Astrophysical Sciences, Princeton University\\
Princeton, NJ 08544, USA}

\author{Volodymyr Takhistov}
\email[]{vtakhist@physics.ucla.edu}
\affiliation{Department of Physics and Astronomy, University of California, Los Angeles\\
Los Angeles, CA 90095-1547, USA}

\date{\today}

\begin{abstract}
Neutron-rich material ejected from neutron star--neutron star (NS--NS) and neutron star--black hole (NS--BH) binary mergers is heated by nuclear processes to temperatures of a few hundred keV, resulting in a population of electron-positron pairs.  Some of the positrons escape from the outer layers of the ejecta.  We show that the population of low-energy positrons produced by NS-NS and NS-BH mergers in the Milky Way can account for the observed 511-keV line from the Galactic center (GC). Moreover, we suggest how positrons and the associated 511-keV emission can be used as tracers of recent mergers. Recent discovery of 511 keV emission from the ultra-faint dwarf galaxy Reticulum II, consistent with a rare NS-NS merger event, provides a smoking-gun signature of our proposal.
\end{abstract}

\maketitle

Recent joint multi-messenger observation of a binary merger event has confirmed that neutron star mergers are a source of both gravitational as well as electromagnetic radiation~\cite{GBM:2017lvd}.~The ejected dense neutron-rich material provides a favorable setting for $r$-process nucleosynthesis, possibly producing heavy elements such as Gold and Uranium~\cite{Kasen:2017sxr, Pian:2017gtc, Drout:2017ijr}, and powers electromagnetic transients known as kilonovae~\cite{Li:1998bw, Metzger:2016pju}.  There is no doubt that NS--NS mergers and NS--BH mergers take place in the Milky Way galaxy as well.  In this {\it Letter} we show that positron production accompanying such mergers can explain the observed 511-keV line from the Galactic center and also serve, along with associated radiation, as a novel tracer of neutron star binary merger events.

The 511-keV line has been consistently observed for several decades~\cite{Johnson:1972,Leventhal:1978} from the Galactic central 
region, with precise
measurements performed by the SPI spectrometer aboard the INTEGRAL satellite~\cite{Knodlseder:2005yq,Siegert:2016ijv}.~The detected flux of the line
in the Galactic bulge component is
$(0.96 \pm 0.07) \times 10^{-3}$ photons cm$^{-2}$ s$^{-1}$~\cite{Siegert:2015knp}.
The signal is consistent with electron-positron annihilation via positronium bound state formation, occurring at a rate of
$\Gamma(e^+e^-\rightarrow \gamma\gamma) \sim 
 10^{50}\, {\rm yr}^{-1}$.~The origin of the positrons remains unknown (for a review see \cite{Prantzos:2010wi}).
Among the possible sources of the positrons are accretion outflows from the GC super-massive black hole Sagittarius (Sgr) A$^{\ast}$ \cite{Totani:2006zx}, pulsar winds \cite{Wang:2005cqa}, X-ray binary/micro-quasar jets~\cite{Guessoum:2006fs,Bandyopadhyay:2008ts}, gamma-ray bursts~\cite{Bertone:2004ek} and radioactive emissions due to nucleosynthesis in massive stars, supernovae, novae and hypernovae~\cite{Prantzos:2010wi, Perets:2014mba,Alexis:2014rba,Milne:2001zs}. More exotic proposals, such as WIMP particle dark matter annihilations~\cite{Boehm:2003bt} and de-excitations~\cite{Finkbeiner:2007kk,Pearce:2015zca} as well as $r$-process nucleosynthesis emission due to neutron star disruptions by primordial black holes~\cite{Fuller:2017uyd} have also been put forward. The dark matter annihilation scenario is already under pressure from cosmological observations~\cite{Frey:2013wh,Wilkinson:2016gsy}.

The source of positrons responsible for the 511-keV line must generate $\sim 10^{50}$ positrons per year.  Furthermore, the positron energies must not exceed 3~MeV if the positrons are to cool and form positronium rather than annihilate in flight~\cite{Beacom:2005qv}.  We will show that NS-NS and NS-BH mergers are capable of producing the requisite numbers of cold positrons. The expanding ejecta is heated by $\beta$-decays and fission to temperatures of a few hundred keV, at which some population of positrons exists in thermal equilibrium.  Some of the positrons will escape from the outer layers of ejecta and produce the observed 511-keV emission line.

\begin{figure*}[ht!]
	\resizebox{\textwidth}{!}{
		\includegraphics{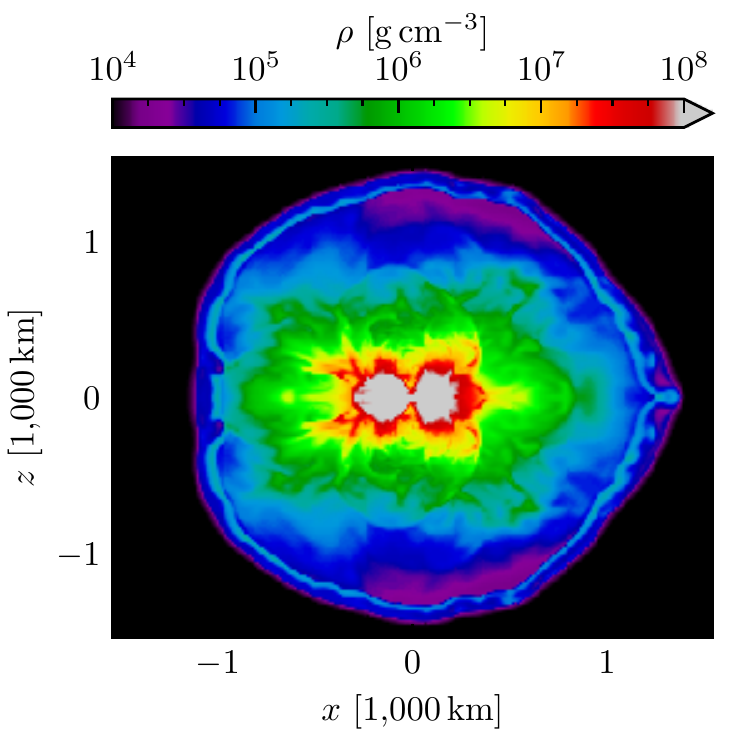}
    	\includegraphics{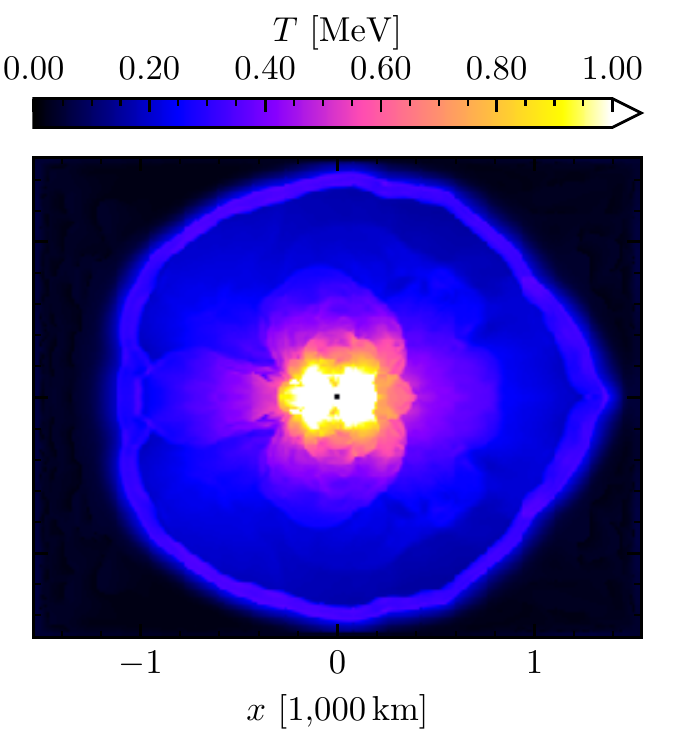}
    	\includegraphics{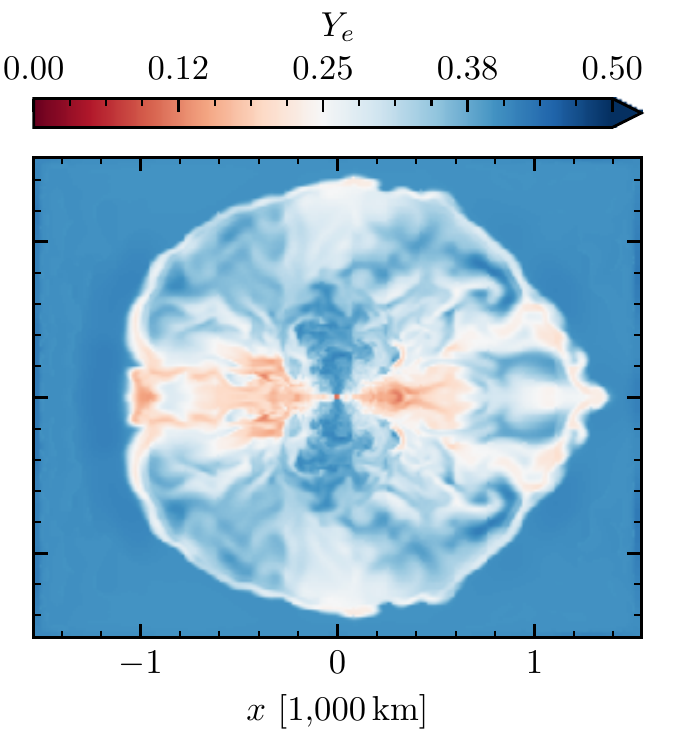}
    }
	\caption{Density (left), temperature (center) and electron fraction (right) profiles of the ejected material at $t = 10$ ms after merger.}
	\label{fig:ejecta} 
\end{figure*}

\begin{figure}[t]
      \includegraphics[width=0.45\textwidth]{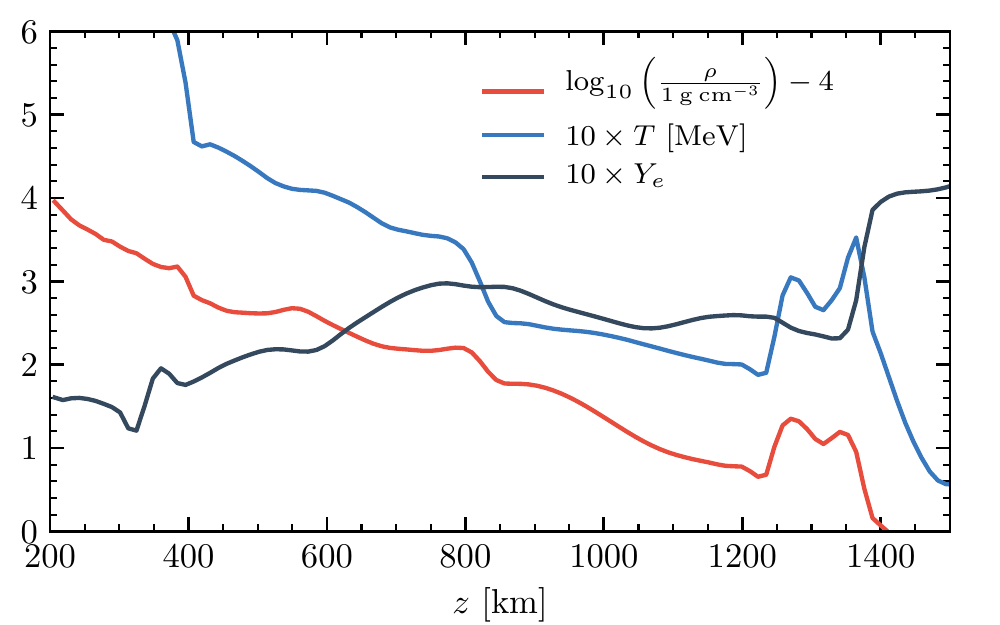}
\caption{1-D profile slices of density (red), temperature (blue) and electron fraction (black) of the ejected material at $t = 10$ ms after merger.}
\label{fig:profiles} 
\end{figure}

As a starting point, we consider the results from state-of-the-art numerical relativity simulations of binary NS mergers, which we have performed by employing the \texttt{WhiskyTHC} code \cite{radice:2012cu, radice:2013hxh, radice:2013xpa}.~The simulations tracked the expansion of the ejecta cloud in the first few milliseconds after the merger and recorded the specific time-dependence of the temperature $T$, electron fraction $Y_e$, as well as density $\rho$ profiles. As a representative example, we consider a binary composed of two $1.35\ M_\odot$ NSs simulated with the microphysical SFHo equation of state \cite{steiner:2012rk}. The effects of weak interactions as well as neutrinos are included using the M0 scheme \cite{radice:2016dwd}.~A detailed account of the simulations is presented in Ref.~\cite{Radice:2018pdn}. The resulting profiles at $t = 10$ ms after the merger are displayed in Fig.~\ref{fig:ejecta}. The associated 1-D profiles along the direction orthogonal to the orbital plane are shown in Fig.~\ref{fig:profiles}.~During the very early stages of the material ejection any inhomogeneities and small-scale structures are rapidly eliminated and the torus-shaped ejecta expand homologously soon after the merger \cite{Perego:2017wtu}, centered around a compact remnant.~Features associated with the expanding shock near the outer ejecta surface layers can be readily observed.~Long term simulations of the average quantities in the dynamical ejecta, including the effects of nuclear process heating, have been studied in Ref.~\cite{Rosswog:2013kqa}. Nucleosynthesis with similar characteristics is also expected from NS-BH mergers \cite{Kawaguchi:2016ana} and in what follows we do not distinguish between NS-NS and NS-BH positron production.

The temperature in the ejecta is always $T \lesssim 1$~MeV. Hence, the resulting positrons are not highly relativistic:  $(\gamma - 1) \simeq (3T/m_e)$, where $\gamma$ is the Lorentz factor. Units of $c = 1$ are assumed throughout. The number density of positrons at a particular temperature follows a Boltzmann distribution: 
\begin{equation} \label{eq:boltzmann}
n(T) = 2 \Big(\dfrac{m_e T}{2 \pi^3}\Big)^{3/2} e^{-m_e/T}~.
\end{equation}
Produced positrons can be confined by magnetic fields~\cite{Barnes:2016umi}, but the confinement cannot be perfect, especially in the presence of random magnetic fields. Hence, a fraction of particles will escape.  While the details related to magnetic field confinement are difficult to analyze without involved simulations, similar studies of the ejecta emission from supernovae (SN) have shown that $\mathcal{O}(10)\%$ of all positrons are expected to escape \cite{Prantzos:2010wi}.

The ability of positrons to penetrate the outer layers of ejecta from radii above $r$ is described by ``optical depth''
\begin{equation}
\tau_e(r) = \int_r^{\infty} ds \rho(s) k_e~,
\end{equation}
where $\rho$ is the density and $k_e$ is the effective electron opacity parameter.~For positrons with energy $E_e \sim \mathcal{O}(1)$~MeV, the average energy loss per unit density is given by the Bethe-Bloch formula as $\langle dE_e/dx \rangle \simeq 1$ MeV cm$^2$/g, resulting in $k_e \simeq 1$ cm$^2$/g. Positron emission can occur from regions that are ``optically thin'', i.e.~where~$\tau_e \lesssim 1$. Since the outer layer density in the initial evolution stages of the ejecta is $\rho \sim 10^4$ g/cm$^3$ (see Fig.~\ref{fig:profiles}), the ejecta is fully opaque to positrons at this time. From analytic models of kilonova \cite{Waxman:2017sqv,Metzger:2016pju}, the density time-dependence approximately follows $\rho \sim t^{-3}$ scaling. Hence, at later times, when the density significantly decreases due to adiabatic expansion, some outer ejecta layers become optically thin to positrons. However, since the required density drop is $\gtrsim5$ orders of magnitude, the accompanying drop in temperature by more than an order of magnitude (see~e.g.~\cite{Rosswog:2013kqa}) would render the late-time positron emission ineffective, in light of Eq.~\eqref{eq:boltzmann}.

Based on physical arguments (e.g.~\cite{Qian:1993dg}), it is generically expected that a thin, low-density, ``atmospheric'' layer of material exists on the outskirts of the ejecta.~In fact, it can be analytically proven~\cite{Munz:1994} that a gas flow (i.e. ejecta) cannot have a shock discontinuity at the interface with vacuum and a rarefaction wave is present instead, implying that such an atmospheric layer with a decreasing density is always there.
Hence, while the ejecta as resolved within initial merger simulations is optically thick to positrons, by continuity, there is some rarefied layer in the outer ``atmosphere'' beyond which the optical depth is less than one. This can be further understood in analogy with a photo-sphere of a star or a neutrino-sphere of a cooling proto-neutron star.  This {\it positron-sphere} is the layer that determines the flux and the mean energy of emitted positrons.   We model the atmosphere as a layer with an exponentially decreasing density, well below the resolution of initial merger simulations (i.e. $\rho \ll 10^4$ g/cm$^3$). Taking the outermost radius resolved in merger simulations as $r_s$, the thin atmospheric layer at $r > r_s$ can be described as \cite{Duan:2006an}
\begin{equation}
\rho_a (r) = \rho_s e^{-(r-r_s)/h}~,
\end{equation}
where $\rho_s = 6.3 \times 10^3$ g/cm$^{3}$, $r_s = 1540$ km have been chosen as the density and radius of the outermost layer from the initial merger simulations at $t = 20$ ms (see Fig.~\ref{fig:profiles}), with $h \simeq 0.056$ km denoting the density normalization parameter as in \cite{Duan:2006an}.
Hence, at any given time there exists an outer layer where the density is low enough to allow for positron emission in the optically thin regime. Schematic representation of the atmospheric layer can be found in Supplemental Material. Solving for $\tau_e = 1$, the radius above which the ejecta is optically thin during the time interval $t_e$ after the merger is given by
\begin{equation}
r_t = r_s + h \log\Big[h k_e \rho_s\Big] +   v_e t_e~,
\end{equation}
where $v_e \simeq 0.8$ is the thermal positron velocity.

Following \cite{Fuller:2017uyd}, we now estimate the contribution to the 511-keV emission line from NS-NS and NS-BH mergers.
The emission surface of the positrons during the time interval $t_e$ after the merger is given by
\begin{equation}
S = 4 \pi r_t^2~.
\end{equation}
We note that the resulting $S$ is not very sensitive to the choice of value for  the density normalization parameter $h$. For temperature we assume $T \simeq 0.1~\text{MeV} \simeq  1.2 \times 10^9$~K, following the initial merger simulations.
Hence, the number of positrons emitted during $t_e\sim1$~s is
\begin{equation}
N_p = n(T) S v_e t_e \simeq 5 \times 10^{58}~,
\end{equation}
Here we have considered the positron emission time interval of $20~{\rm ms} \lesssim t_e \lesssim 1~{\rm s}$ and the range of the corresponding emission radius as $10^3~{\rm km} \lesssim r_t \lesssim 10^5~{\rm km}$. The lower values match the initial merger simulations, while the upper values correspond to the time interval during which the temperature has remained approximately constant throughout the emission. As the ejecta expands, it can be shown (see Supplemental Material~\cite{suppmat}) that adiabatic and radiative cooling cause the temperature to decrease linearly with time.
With a starting time of $t_0 \sim \mathcal{O}(1)$ ms, for $t_e \sim \mathcal{O}(1)$ s the cooling is significant. However, this neglects nuclear heating. When the full network of nuclear processes is taken into account, long-term ejecta evolution simulations demonstrate that the additional heating  raises the average ejecta temperature for a few seconds~\cite{Rosswog:2013kqa}. Hence, the assumption of approximately constant temperature during the time interval $t_e$ is justified.

The cumulative Galactic merger rate of NS-NS and NS-BH binaries, as inferred from Advanced LIGO's first observation run, is around few$\times 100$ Myr$^{-1}$~\cite{Abbott:2016ymx}. We note that the merger rate is highly uncertain and a better understanding of therates in the bulge and in the disk could allow one to use the signal morphology to gain further insights into the origin of 511-keV emission. We conservatively take the merger rate to be $R_{MW} \simeq (10^{-2} - 10^2)$ Myr$^{-1}$ \cite{Mapelli:2018wys, Chruslinska:2017odi}. The resulting average positron emission rate is then
\begin{equation}
\Gamma = N_p R_{MW} \simeq 5 \times 10^{50-54}~\text{yr}^{-1}~.
\end{equation}
We note that the value of $\Gamma$ is subject to various astrophysical uncertainties, such as the binary merger rate, geometry of the outflows, magnetic fields in the ejecta, etc. 

The expected morphology of the 511 keV signal in the Galactic bulge, disk and halo regions can be heuristically understood as follows. The total distance that positrons with energy $E_e \sim \mathcal{O}(1)$~MeV will propagate through diffusion within the interstellar medium from their birth sites, including collisional as well as collisionless plasma transport regimes, is $r_d \sim \mathcal{O}(100)$ pc \cite{Jean:2009zj}.~The bulge component associated with the 511 keV excess extends up to $\sim1.5$ kpc. Since the binary merger timescales of $1/R_{MW} \simeq 2 \times 10^4~\text{yrs}$ are far below the diffusion timescales of $\tau_d \simeq 10^{7-8}$ yrs \cite{Bertone:2004ek}, the bulge can be fully populated by the positrons as desired.~The 511-keV line has also been observed from the Galactic disk~\cite{Siegert:2015knp}, which indicates that the origin could be related to the stellar population.~This favors a binary merger origin over some alternatives, such as those involving dark matter annihilation or decay.~While nucleosynthesis from type-Ia supernovae has been also suggested as a source of considerable 511 keV signal~\cite{Prantzos:2010wi}, it is challenging for this proposal to address the observed disk signal contribution due to the Galactic Center-oriented distribution of type-Ia supernovae. On the other hand, binary mergers are expected to have a sizable $\mathcal{O}(20)\%$ non-bulge component due to binary kicks \cite{Berger:2013jza}, which in our scenario results in a non-negligible signal contribution from the disk, as observed. In the halo, away from the disk, the gas density and the magnetic fields are small and the positrons become de-localized before annihilating. Since the positron energy losses depend sensitively on the gas density and the structure of the magnetic fields~\cite{Bertone:2004ek}, we do not expect a sizable signal coming from that region. We further note that we do not expect a significant signal contribution from globular clusters. While globular clusters contain a large stellar density, NS kicks ensure that the NS-binary merger rate in globular clusters, which is highly uncertain, is small and contributes only at the $< \mathcal{O}(1)\%$ level to the overall Galactic NS-binary merger rate~\cite{Tsang:2013mca}.

Positrons and the associated 511 keV radiation can serve as novel tracers of recent binary neutron star mergers. In particular, positrons can provide direct detection of a merger event within the local binary population. Further, positrons born in hot outflows, with plasma temperatures $\gtrsim 10^7$ K, may not annihilate until the plasma has cooled and could be advected to distances up to few kpc \cite{Jean:2009zj,Churazov:2010wy}. Neutron star mergers are expected to occur not only in the Galactic center, but also in globular clusters. Since the two nearest globular clusters are located 2.2 kpc (NGC 6121/M4) kpc and 2.4 kpc (NGC 6397) away, they serve as potential candidate sites for merger positron detection.
The positron population in a certain Galactic region can also be indicative of the associated binary merger history. 

Analogously, gamma-rays associated with positron annihilations allow for an indirect detection of a neutron star merger event. In particular, 1 out of $10^{2}{-}10^{3}$ SN-remnant-like objects could in fact be a binary NS merger remnant, according to current estimates for SN and NS merger rates \cite{Li:2010kd, Adams:2013ana, TheLIGOScientific:2017qsa, Chruslinska:2017odi}. Since the associated positron luminosity from the merger is several orders above that of SN, we predict that observation of a bright emission hot-spot in the 511 keV radiation spectrum will allow discrimination between the merger and SN remnants. 

A spectroscopic survey \cite{Ji:2015wzg} of ultra-faint dwarf galaxies (UFDs) suggests that r-process nucleosynthesis is a rare event, a conclusion consistent with binary neutron star mergers as the origin of this material. Out of 10 studied UFDs, only Reticulum II showed r-process enhancement. While UFDs have shallow gravitational potential wells and retainment of NS binaries might appear difficult due to natal kicks, recent analyses suggest that a significant binary fraction will not escape from the host galaxy~\cite{Beniamini:2017qqy}.~Tantalizingly, in INTEGRAL gamma-ray observations \cite{Siegert:2016ijv}, Reticulum II was also the only dwarf galaxy to show a significant 511 keV photon flux. This concordance of the nucleosynthesis and gamma-ray observations is naturally expected in the context of binary merger origin for positrons, as suggested here.

We have demonstrated that the Galactic positron production from binary mergers is consistent in energetics, as well as in rate, with the observed 511-keV GC emission line. Positron production can be used as a tracer of merger activity in combination with other multi-messenger signals. Recent discovery of 511 keV emission from dwarf galaxies provides a smoking gun for our proposal.

\section{Acknowledgments}
We would like to thank Kenta Hotokezaka for discussion. The work of G.M.F. was supported in part by National Science Foundation Grant No. PHY-1614864
and the NSF N3AS Hub, NSF Grant No. PHY-1630782 and Heising-Simons Foundation Grant No. 2017-228.
G.M.F. also acknowledges Department of Energy Scientific Discovery through Advanced Computing (SciDAC-4) grant register No. SN60152, award No. DE-SC0018297.
The work of A.K. and V.T. was supported by the U.S. Department of Energy Grant No. DE-SC0009937. A.K. was also supported by the World Premier International Research Center Initiative (WPI), MEXT, Japan. 
D.R. gratefully acknowledges support from a Frank and Peggy Taplin Membership at the Institute for Advanced Study and the Max-Planck/Princeton Center (MPPC) for Plasma Physics (NSF PHY-1804048).
Computations were performed on the supercomputers Bridges, Comet, and Stampede (NSF XSEDE allocation TG-PHY160025), and on NSF/NCSA Blue Waters (NSF PRAC ACI-1440083 and NSF PRAC AWD-1811236).

\bibliography{nsnspos}

\begin{thebibliography}{54}%
\makeatletter
\providecommand \@ifxundefined [1]{%
 \@ifx{#1\undefined}
}%
\providecommand \@ifnum [1]{%
 \ifnum #1\expandafter \@firstoftwo
 \else \expandafter \@secondoftwo
 \fi
}%
\providecommand \@ifx [1]{%
 \ifx #1\expandafter \@firstoftwo
 \else \expandafter \@secondoftwo
 \fi
}%
\providecommand \natexlab [1]{#1}%
\providecommand \enquote  [1]{``#1''}%
\providecommand \bibnamefont  [1]{#1}%
\providecommand \bibfnamefont [1]{#1}%
\providecommand \citenamefont [1]{#1}%
\providecommand \href@noop [0]{\@secondoftwo}%
\providecommand \href [0]{\begingroup \@sanitize@url \@href}%
\providecommand \@href[1]{\@@startlink{#1}\@@href}%
\providecommand \@@href[1]{\endgroup#1\@@endlink}%
\providecommand \@sanitize@url [0]{\catcode `\\12\catcode `\$12\catcode
  `\&12\catcode `\#12\catcode `\^12\catcode `\_12\catcode `\%12\relax}%
\providecommand \@@startlink[1]{}%
\providecommand \@@endlink[0]{}%
\providecommand \url  [0]{\begingroup\@sanitize@url \@url }%
\providecommand \@url [1]{\endgroup\@href {#1}{\urlprefix }}%
\providecommand \urlprefix  [0]{URL }%
\providecommand \Eprint [0]{\href }%
\providecommand \doibase [0]{http://dx.doi.org/}%
\providecommand \selectlanguage [0]{\@gobble}%
\providecommand \bibinfo  [0]{\@secondoftwo}%
\providecommand \bibfield  [0]{\@secondoftwo}%
\providecommand \translation [1]{[#1]}%
\providecommand \BibitemOpen [0]{}%
\providecommand \bibitemStop [0]{}%
\providecommand \bibitemNoStop [0]{.\EOS\space}%
\providecommand \EOS [0]{\spacefactor3000\relax}%
\providecommand \BibitemShut  [1]{\csname bibitem#1\endcsname}%
\let\auto@bib@innerbib\@empty
\bibitem [{\citenamefont {Abbott}\ \emph
  {et~al.}(2017{\natexlab{a}})\citenamefont {Abbott} \emph
  {et~al.}}]{GBM:2017lvd}%
  \BibitemOpen
  \bibfield  {author} {\bibinfo {author} {\bibfnamefont {B.~P.}\ \bibnamefont
  {Abbott}} \emph {et~al.} (\bibinfo {collaboration} {GROND, SALT Group,
  OzGrav, DFN, INTEGRAL, Virgo, Insight-Hxmt, MAXI Team, Fermi-LAT, J-GEM,
  RATIR, IceCube, CAASTRO, LWA, ePESSTO, GRAWITA, RIMAS, SKA South
  Africa/MeerKAT, H.E.S.S., 1M2H Team, IKI-GW Follow-up, Fermi GBM, Pi of Sky,
  DWF (Deeper Wider Faster Program), Dark Energy Survey, MASTER, AstroSat
  Cadmium Zinc Telluride Imager Team, Swift, Pierre Auger, ASKAP, VINROUGE,
  JAGWAR, Chandra Team at McGill University, TTU-NRAO, GROWTH, AGILE Team, MWA,
  ATCA, AST3, TOROS, Pan-STARRS, NuSTAR, ATLAS Telescopes, BOOTES, CaltechNRAO,
  LIGO Scientific, High Time Resolution Universe Survey, Nordic Optical
  Telescope, Las Cumbres Observatory Group, TZAC Consortium, LOFAR, IPN, DLT40,
  Texas Tech University, HAWC, ANTARES, KU, Dark Energy Camera GW-EM, CALET,
  Euro VLBI Team, ALMA}),\ }\href {\doibase 10.3847/2041-8213/aa91c9}
  {\bibfield  {journal} {\bibinfo  {journal} {Astrophys. J.}\ }\textbf
  {\bibinfo {volume} {848}},\ \bibinfo {pages} {L12} (\bibinfo {year}
  {2017}{\natexlab{a}})},\ \Eprint {http://arxiv.org/abs/1710.05833}
  {arXiv:1710.05833 [astro-ph.HE]} \BibitemShut {NoStop}%
\bibitem [{\citenamefont {Kasen}\ \emph {et~al.}(2017)\citenamefont {Kasen},
  \citenamefont {Metzger}, \citenamefont {Barnes}, \citenamefont {Quataert},\
  and\ \citenamefont {Ramirez-Ruiz}}]{Kasen:2017sxr}%
  \BibitemOpen
  \bibfield  {author} {\bibinfo {author} {\bibfnamefont {D.}~\bibnamefont
  {Kasen}}, \bibinfo {author} {\bibfnamefont {B.}~\bibnamefont {Metzger}},
  \bibinfo {author} {\bibfnamefont {J.}~\bibnamefont {Barnes}}, \bibinfo
  {author} {\bibfnamefont {E.}~\bibnamefont {Quataert}}, \ and\ \bibinfo
  {author} {\bibfnamefont {E.}~\bibnamefont {Ramirez-Ruiz}},\ }\href {\doibase
  10.1038/nature24453} {\bibfield  {journal} {\bibinfo  {journal} {Nature}\ }
  (\bibinfo {year} {2017}),\ 10.1038/nature24453},\ \bibinfo {note}
  {[Nature551,80(2017)]},\ \Eprint {http://arxiv.org/abs/1710.05463}
  {arXiv:1710.05463 [astro-ph.HE]} \BibitemShut {NoStop}%
\bibitem [{\citenamefont {Pian}\ \emph {et~al.}(2017)\citenamefont {Pian} \emph
  {et~al.}}]{Pian:2017gtc}%
  \BibitemOpen
  \bibfield  {author} {\bibinfo {author} {\bibfnamefont {E.}~\bibnamefont
  {Pian}} \emph {et~al.},\ }\href {\doibase 10.1038/nature24298} {\bibfield
  {journal} {\bibinfo  {journal} {Nature}\ }\textbf {\bibinfo {volume} {551}},\
  \bibinfo {pages} {67} (\bibinfo {year} {2017})},\ \Eprint
  {http://arxiv.org/abs/1710.05858} {arXiv:1710.05858 [astro-ph.HE]}
  \BibitemShut {NoStop}%
\bibitem [{\citenamefont {Drout}\ \emph {et~al.}(2017)\citenamefont {Drout}
  \emph {et~al.}}]{Drout:2017ijr}%
  \BibitemOpen
  \bibfield  {author} {\bibinfo {author} {\bibfnamefont {M.~R.}\ \bibnamefont
  {Drout}} \emph {et~al.},\ }\href {\doibase 10.1126/science.aaq0049}
  {\bibfield  {journal} {\bibinfo  {journal} {Science}\ }\textbf {\bibinfo
  {volume} {358}},\ \bibinfo {pages} {1570} (\bibinfo {year} {2017})},\ \Eprint
  {http://arxiv.org/abs/1710.05443} {arXiv:1710.05443 [astro-ph.HE]}
  \BibitemShut {NoStop}%
\bibitem [{\citenamefont {Li}\ and\ \citenamefont
  {Paczynski}(1998)}]{Li:1998bw}%
  \BibitemOpen
  \bibfield  {author} {\bibinfo {author} {\bibfnamefont {L.-X.}\ \bibnamefont
  {Li}}\ and\ \bibinfo {author} {\bibfnamefont {B.}~\bibnamefont {Paczynski}},\
  }\href {\doibase 10.1086/311680} {\bibfield  {journal} {\bibinfo  {journal}
  {Astrophys. J.}\ }\textbf {\bibinfo {volume} {507}},\ \bibinfo {pages} {L59}
  (\bibinfo {year} {1998})},\ \Eprint {http://arxiv.org/abs/astro-ph/9807272}
  {arXiv:astro-ph/9807272 [astro-ph]} \BibitemShut {NoStop}%
\bibitem [{\citenamefont {Metzger}(2017)}]{Metzger:2016pju}%
  \BibitemOpen
  \bibfield  {author} {\bibinfo {author} {\bibfnamefont {B.~D.}\ \bibnamefont
  {Metzger}},\ }\href {\doibase 10.1007/s41114-017-0006-z} {\bibfield
  {journal} {\bibinfo  {journal} {Living Rev. Rel.}\ }\textbf {\bibinfo
  {volume} {20}},\ \bibinfo {pages} {3} (\bibinfo {year} {2017})},\ \Eprint
  {http://arxiv.org/abs/1610.09381} {arXiv:1610.09381 [astro-ph.HE]}
  \BibitemShut {NoStop}%
\bibitem [{\citenamefont {{Johnson}}\ \emph {et~al.}(1972)\citenamefont
  {{Johnson}}, \citenamefont {{Harnden}},\ and\ \citenamefont
  {{Haymes}}}]{Johnson:1972}%
  \BibitemOpen
  \bibfield  {author} {\bibinfo {author} {\bibfnamefont {W.~N.}\ \bibnamefont
  {{Johnson}}, \bibfnamefont {III}}, \bibinfo {author} {\bibfnamefont {F.~R.}\
  \bibnamefont {{Harnden}}, \bibfnamefont {Jr.}}, \ and\ \bibinfo {author}
  {\bibfnamefont {R.~C.}\ \bibnamefont {{Haymes}}},\ }\href {\doibase
  10.1086/180878} {\bibfield  {journal} {\bibinfo  {journal} {Astrophys. J.}\
  }\textbf {\bibinfo {volume} {172}},\ \bibinfo {pages} {L1} (\bibinfo {year}
  {1972})}\BibitemShut {NoStop}%
\bibitem [{\citenamefont {{Leventhal}}\ \emph {et~al.}(1978)\citenamefont
  {{Leventhal}}, \citenamefont {{MacCallum}},\ and\ \citenamefont
  {{Stang}}}]{Leventhal:1978}%
  \BibitemOpen
  \bibfield  {author} {\bibinfo {author} {\bibfnamefont {M.}~\bibnamefont
  {{Leventhal}}}, \bibinfo {author} {\bibfnamefont {C.~J.}\ \bibnamefont
  {{MacCallum}}}, \ and\ \bibinfo {author} {\bibfnamefont {P.~D.}\ \bibnamefont
  {{Stang}}},\ }\href {\doibase 10.1086/182782} {\bibfield  {journal} {\bibinfo
   {journal} {Astrophys. J.}\ }\textbf {\bibinfo {volume} {225}},\ \bibinfo
  {pages} {L11} (\bibinfo {year} {1978})}\BibitemShut {NoStop}%
\bibitem [{\citenamefont {Knodlseder}\ \emph {et~al.}(2005)\citenamefont
  {Knodlseder} \emph {et~al.}}]{Knodlseder:2005yq}%
  \BibitemOpen
  \bibfield  {author} {\bibinfo {author} {\bibfnamefont {J.}~\bibnamefont
  {Knodlseder}} \emph {et~al.},\ }\href {\doibase 10.1051/0004-6361:20042063}
  {\bibfield  {journal} {\bibinfo  {journal} {Astron. Astrophys.}\ }\textbf
  {\bibinfo {volume} {441}},\ \bibinfo {pages} {513} (\bibinfo {year}
  {2005})},\ \Eprint {http://arxiv.org/abs/astro-ph/0506026}
  {arXiv:astro-ph/0506026 [astro-ph]} \BibitemShut {NoStop}%
\bibitem [{\citenamefont {Siegert}\ \emph
  {et~al.}(2016{\natexlab{a}})\citenamefont {Siegert}, \citenamefont {Diehl},
  \citenamefont {Vincent}, \citenamefont {Guglielmetti}, \citenamefont
  {Krause},\ and\ \citenamefont {Boehm}}]{Siegert:2016ijv}%
  \BibitemOpen
  \bibfield  {author} {\bibinfo {author} {\bibfnamefont {T.}~\bibnamefont
  {Siegert}}, \bibinfo {author} {\bibfnamefont {R.}~\bibnamefont {Diehl}},
  \bibinfo {author} {\bibfnamefont {A.~C.}\ \bibnamefont {Vincent}}, \bibinfo
  {author} {\bibfnamefont {F.}~\bibnamefont {Guglielmetti}}, \bibinfo {author}
  {\bibfnamefont {M.~G.~H.}\ \bibnamefont {Krause}}, \ and\ \bibinfo {author}
  {\bibfnamefont {C.}~\bibnamefont {Boehm}},\ }\href {\doibase
  10.1051/0004-6361/201629136} {\bibfield  {journal} {\bibinfo  {journal}
  {Astron. Astrophys.}\ }\textbf {\bibinfo {volume} {595}},\ \bibinfo {pages}
  {A25} (\bibinfo {year} {2016}{\natexlab{a}})},\ \Eprint
  {http://arxiv.org/abs/1608.00393} {arXiv:1608.00393 [astro-ph.HE]}
  \BibitemShut {NoStop}%
\bibitem [{\citenamefont {Siegert}\ \emph
  {et~al.}(2016{\natexlab{b}})\citenamefont {Siegert}, \citenamefont {Diehl},
  \citenamefont {Khachatryan}, \citenamefont {Krause}, \citenamefont
  {Guglielmetti}, \citenamefont {Greiner}, \citenamefont {Strong},\ and\
  \citenamefont {Zhang}}]{Siegert:2015knp}%
  \BibitemOpen
  \bibfield  {author} {\bibinfo {author} {\bibfnamefont {T.}~\bibnamefont
  {Siegert}}, \bibinfo {author} {\bibfnamefont {R.}~\bibnamefont {Diehl}},
  \bibinfo {author} {\bibfnamefont {G.}~\bibnamefont {Khachatryan}}, \bibinfo
  {author} {\bibfnamefont {M.~G.~H.}\ \bibnamefont {Krause}}, \bibinfo {author}
  {\bibfnamefont {F.}~\bibnamefont {Guglielmetti}}, \bibinfo {author}
  {\bibfnamefont {J.}~\bibnamefont {Greiner}}, \bibinfo {author} {\bibfnamefont
  {A.~W.}\ \bibnamefont {Strong}}, \ and\ \bibinfo {author} {\bibfnamefont
  {X.}~\bibnamefont {Zhang}},\ }\href {\doibase 10.1051/0004-6361/201527510}
  {\bibfield  {journal} {\bibinfo  {journal} {Astron. Astrophys.}\ }\textbf
  {\bibinfo {volume} {586}},\ \bibinfo {pages} {A84} (\bibinfo {year}
  {2016}{\natexlab{b}})},\ \Eprint {http://arxiv.org/abs/1512.00325}
  {arXiv:1512.00325 [astro-ph.HE]} \BibitemShut {NoStop}%
\bibitem [{\citenamefont {Prantzos}\ \emph {et~al.}(2011)\citenamefont
  {Prantzos} \emph {et~al.}}]{Prantzos:2010wi}%
  \BibitemOpen
  \bibfield  {author} {\bibinfo {author} {\bibfnamefont {N.}~\bibnamefont
  {Prantzos}} \emph {et~al.},\ }\href {\doibase 10.1103/RevModPhys.83.1001}
  {\bibfield  {journal} {\bibinfo  {journal} {Rev. Mod. Phys.}\ }\textbf
  {\bibinfo {volume} {83}},\ \bibinfo {pages} {1001} (\bibinfo {year}
  {2011})},\ \Eprint {http://arxiv.org/abs/1009.4620} {arXiv:1009.4620
  [astro-ph.HE]} \BibitemShut {NoStop}%
\bibitem [{\citenamefont {Totani}(2006)}]{Totani:2006zx}%
  \BibitemOpen
  \bibfield  {author} {\bibinfo {author} {\bibfnamefont {T.}~\bibnamefont
  {Totani}},\ }\href {\doibase 10.1093/pasj/58.6.965} {\bibfield  {journal}
  {\bibinfo  {journal} {Publ. Astron. Soc. Jap.}\ }\textbf {\bibinfo {volume}
  {58}},\ \bibinfo {pages} {965} (\bibinfo {year} {2006})},\ \Eprint
  {http://arxiv.org/abs/astro-ph/0607414} {arXiv:astro-ph/0607414 [astro-ph]}
  \BibitemShut {NoStop}%
\bibitem [{\citenamefont {Wang}\ \emph {et~al.}(2006)\citenamefont {Wang},
  \citenamefont {Pun},\ and\ \citenamefont {Cheng}}]{Wang:2005cqa}%
  \BibitemOpen
  \bibfield  {author} {\bibinfo {author} {\bibfnamefont {W.}~\bibnamefont
  {Wang}}, \bibinfo {author} {\bibfnamefont {C.~S.~J.}\ \bibnamefont {Pun}}, \
  and\ \bibinfo {author} {\bibfnamefont {K.~S.}\ \bibnamefont {Cheng}},\ }\href
  {\doibase 10.1051/0004-6361:20053559} {\bibfield  {journal} {\bibinfo
  {journal} {Astron. Astrophys.}\ }\textbf {\bibinfo {volume} {446}},\ \bibinfo
  {pages} {943} (\bibinfo {year} {2006})},\ \Eprint
  {http://arxiv.org/abs/astro-ph/0509760} {arXiv:astro-ph/0509760 [astro-ph]}
  \BibitemShut {NoStop}%
\bibitem [{\citenamefont {Guessoum}\ \emph {et~al.}(2006)\citenamefont
  {Guessoum}, \citenamefont {Jean},\ and\ \citenamefont
  {Prantzos}}]{Guessoum:2006fs}%
  \BibitemOpen
  \bibfield  {author} {\bibinfo {author} {\bibfnamefont {N.}~\bibnamefont
  {Guessoum}}, \bibinfo {author} {\bibfnamefont {P.}~\bibnamefont {Jean}}, \
  and\ \bibinfo {author} {\bibfnamefont {N.}~\bibnamefont {Prantzos}},\ }\href
  {\doibase 10.1051/0004-6361:20065240} {\bibfield  {journal} {\bibinfo
  {journal} {Astron. Astrophys.}\ } (\bibinfo {year} {2006}),\
  10.1051/0004-6361:20065240},\ \bibinfo {note} {[Astron.
  Astrophys.457,753(2006)]},\ \Eprint {http://arxiv.org/abs/astro-ph/0607296}
  {arXiv:astro-ph/0607296 [astro-ph]} \BibitemShut {NoStop}%
\bibitem [{\citenamefont {Bandyopadhyay}\ \emph {et~al.}(2009)\citenamefont
  {Bandyopadhyay}, \citenamefont {Silk}, \citenamefont {Taylor},\ and\
  \citenamefont {Maccarone}}]{Bandyopadhyay:2008ts}%
  \BibitemOpen
  \bibfield  {author} {\bibinfo {author} {\bibfnamefont {R.~M.}\ \bibnamefont
  {Bandyopadhyay}}, \bibinfo {author} {\bibfnamefont {J.}~\bibnamefont {Silk}},
  \bibinfo {author} {\bibfnamefont {J.~E.}\ \bibnamefont {Taylor}}, \ and\
  \bibinfo {author} {\bibfnamefont {T.~J.}\ \bibnamefont {Maccarone}},\ }\href
  {\doibase 10.1111/j.1365-2966.2008.14113.x} {\bibfield  {journal} {\bibinfo
  {journal} {Mon. Not. Roy. Astron. Soc.}\ }\textbf {\bibinfo {volume} {392}},\
  \bibinfo {pages} {1115} (\bibinfo {year} {2009})},\ \Eprint
  {http://arxiv.org/abs/0810.3674} {arXiv:0810.3674 [astro-ph]} \BibitemShut
  {NoStop}%
\bibitem [{\citenamefont {Bertone}\ \emph {et~al.}(2006)\citenamefont
  {Bertone}, \citenamefont {Kusenko}, \citenamefont {Palomares-Ruiz},
  \citenamefont {Pascoli},\ and\ \citenamefont {Semikoz}}]{Bertone:2004ek}%
  \BibitemOpen
  \bibfield  {author} {\bibinfo {author} {\bibfnamefont {G.}~\bibnamefont
  {Bertone}}, \bibinfo {author} {\bibfnamefont {A.}~\bibnamefont {Kusenko}},
  \bibinfo {author} {\bibfnamefont {S.}~\bibnamefont {Palomares-Ruiz}},
  \bibinfo {author} {\bibfnamefont {S.}~\bibnamefont {Pascoli}}, \ and\
  \bibinfo {author} {\bibfnamefont {D.}~\bibnamefont {Semikoz}},\ }\href
  {\doibase 10.1016/j.physletb.2006.03.022} {\bibfield  {journal} {\bibinfo
  {journal} {Phys. Lett.}\ }\textbf {\bibinfo {volume} {B636}},\ \bibinfo
  {pages} {20} (\bibinfo {year} {2006})},\ \Eprint
  {http://arxiv.org/abs/astro-ph/0405005} {arXiv:astro-ph/0405005 [astro-ph]}
  \BibitemShut {NoStop}%
\bibitem [{\citenamefont {Perets}(2014)}]{Perets:2014mba}%
  \BibitemOpen
  \bibfield  {author} {\bibinfo {author} {\bibfnamefont {H.~B.}\ \bibnamefont
  {Perets}},\ }\href@noop {} {\  (\bibinfo {year} {2014})},\ \Eprint
  {http://arxiv.org/abs/1407.2254} {arXiv:1407.2254 [astro-ph.HE]} \BibitemShut
  {NoStop}%
\bibitem [{\citenamefont {Alexis}\ \emph {et~al.}(2014)\citenamefont {Alexis},
  \citenamefont {Jean}, \citenamefont {Martin},\ and\ \citenamefont
  {Ferriere}}]{Alexis:2014rba}%
  \BibitemOpen
  \bibfield  {author} {\bibinfo {author} {\bibfnamefont {A.}~\bibnamefont
  {Alexis}}, \bibinfo {author} {\bibfnamefont {P.}~\bibnamefont {Jean}},
  \bibinfo {author} {\bibfnamefont {P.}~\bibnamefont {Martin}}, \ and\ \bibinfo
  {author} {\bibfnamefont {K.}~\bibnamefont {Ferriere}},\ }\href {\doibase
  10.1051/0004-6361/201322393} {\bibfield  {journal} {\bibinfo  {journal}
  {Astron. Astrophys.}\ }\textbf {\bibinfo {volume} {564}},\ \bibinfo {pages}
  {A108} (\bibinfo {year} {2014})},\ \Eprint {http://arxiv.org/abs/1402.6110}
  {arXiv:1402.6110 [astro-ph.HE]} \BibitemShut {NoStop}%
\bibitem [{\citenamefont {Milne}\ \emph {et~al.}(2002)\citenamefont {Milne},
  \citenamefont {Kurfess}, \citenamefont {Kinzer},\ and\ \citenamefont
  {Leising}}]{Milne:2001zs}%
  \BibitemOpen
  \bibfield  {author} {\bibinfo {author} {\bibfnamefont {P.~A.}\ \bibnamefont
  {Milne}}, \bibinfo {author} {\bibfnamefont {J.~D.}\ \bibnamefont {Kurfess}},
  \bibinfo {author} {\bibfnamefont {R.~L.}\ \bibnamefont {Kinzer}}, \ and\
  \bibinfo {author} {\bibfnamefont {M.~D.}\ \bibnamefont {Leising}},\ }\href
  {\doibase 10.1016/S1387-6473(02)00200-2} {\bibfield  {journal} {\bibinfo
  {journal} {New Astron. Rev.}\ }\textbf {\bibinfo {volume} {46}},\ \bibinfo
  {pages} {553} (\bibinfo {year} {2002})},\ \Eprint
  {http://arxiv.org/abs/astro-ph/0110442} {arXiv:astro-ph/0110442 [astro-ph]}
  \BibitemShut {NoStop}%
\bibitem [{\citenamefont {Boehm}\ \emph {et~al.}(2004)\citenamefont {Boehm},
  \citenamefont {Hooper}, \citenamefont {Silk}, \citenamefont {Casse},\ and\
  \citenamefont {Paul}}]{Boehm:2003bt}%
  \BibitemOpen
  \bibfield  {author} {\bibinfo {author} {\bibfnamefont {C.}~\bibnamefont
  {Boehm}}, \bibinfo {author} {\bibfnamefont {D.}~\bibnamefont {Hooper}},
  \bibinfo {author} {\bibfnamefont {J.}~\bibnamefont {Silk}}, \bibinfo {author}
  {\bibfnamefont {M.}~\bibnamefont {Casse}}, \ and\ \bibinfo {author}
  {\bibfnamefont {J.}~\bibnamefont {Paul}},\ }\href {\doibase
  10.1103/PhysRevLett.92.101301} {\bibfield  {journal} {\bibinfo  {journal}
  {Phys. Rev. Lett.}\ }\textbf {\bibinfo {volume} {92}},\ \bibinfo {pages}
  {101301} (\bibinfo {year} {2004})},\ \Eprint
  {http://arxiv.org/abs/astro-ph/0309686} {arXiv:astro-ph/0309686 [astro-ph]}
  \BibitemShut {NoStop}%
\bibitem [{\citenamefont {Finkbeiner}\ and\ \citenamefont
  {Weiner}(2007)}]{Finkbeiner:2007kk}%
  \BibitemOpen
  \bibfield  {author} {\bibinfo {author} {\bibfnamefont {D.~P.}\ \bibnamefont
  {Finkbeiner}}\ and\ \bibinfo {author} {\bibfnamefont {N.}~\bibnamefont
  {Weiner}},\ }\href {\doibase 10.1103/PhysRevD.76.083519} {\bibfield
  {journal} {\bibinfo  {journal} {Phys. Rev.}\ }\textbf {\bibinfo {volume}
  {D76}},\ \bibinfo {pages} {083519} (\bibinfo {year} {2007})},\ \Eprint
  {http://arxiv.org/abs/astro-ph/0702587} {arXiv:astro-ph/0702587 [astro-ph]}
  \BibitemShut {NoStop}%
\bibitem [{\citenamefont {Pearce}\ \emph {et~al.}(2015)\citenamefont {Pearce},
  \citenamefont {Petraki},\ and\ \citenamefont {Kusenko}}]{Pearce:2015zca}%
  \BibitemOpen
  \bibfield  {author} {\bibinfo {author} {\bibfnamefont {L.}~\bibnamefont
  {Pearce}}, \bibinfo {author} {\bibfnamefont {K.}~\bibnamefont {Petraki}}, \
  and\ \bibinfo {author} {\bibfnamefont {A.}~\bibnamefont {Kusenko}},\ }\href
  {\doibase 10.1103/PhysRevD.91.083532} {\bibfield  {journal} {\bibinfo
  {journal} {Phys. Rev.}\ }\textbf {\bibinfo {volume} {D91}},\ \bibinfo {pages}
  {083532} (\bibinfo {year} {2015})},\ \Eprint
  {http://arxiv.org/abs/1502.01755} {arXiv:1502.01755 [hep-ph]} \BibitemShut
  {NoStop}%
\bibitem [{\citenamefont {Fuller}\ \emph {et~al.}(2017)\citenamefont {Fuller},
  \citenamefont {Kusenko},\ and\ \citenamefont {Takhistov}}]{Fuller:2017uyd}%
  \BibitemOpen
  \bibfield  {author} {\bibinfo {author} {\bibfnamefont {G.~M.}\ \bibnamefont
  {Fuller}}, \bibinfo {author} {\bibfnamefont {A.}~\bibnamefont {Kusenko}}, \
  and\ \bibinfo {author} {\bibfnamefont {V.}~\bibnamefont {Takhistov}},\ }\href
  {\doibase 10.1103/PhysRevLett.119.061101} {\bibfield  {journal} {\bibinfo
  {journal} {Phys. Rev. Lett.}\ }\textbf {\bibinfo {volume} {119}},\ \bibinfo
  {pages} {061101} (\bibinfo {year} {2017})},\ \Eprint
  {http://arxiv.org/abs/1704.01129} {arXiv:1704.01129 [astro-ph.HE]}
  \BibitemShut {NoStop}%
\bibitem [{\citenamefont {Frey}\ and\ \citenamefont
  {Reid}(2013)}]{Frey:2013wh}%
  \BibitemOpen
  \bibfield  {author} {\bibinfo {author} {\bibfnamefont {A.~R.}\ \bibnamefont
  {Frey}}\ and\ \bibinfo {author} {\bibfnamefont {N.~B.}\ \bibnamefont
  {Reid}},\ }\href {\doibase 10.1103/PhysRevD.87.103508} {\bibfield  {journal}
  {\bibinfo  {journal} {Phys. Rev.}\ }\textbf {\bibinfo {volume} {D87}},\
  \bibinfo {pages} {103508} (\bibinfo {year} {2013})},\ \Eprint
  {http://arxiv.org/abs/1301.0819} {arXiv:1301.0819 [hep-ph]} \BibitemShut
  {NoStop}%
\bibitem [{\citenamefont {Wilkinson}\ \emph {et~al.}(2016)\citenamefont
  {Wilkinson}, \citenamefont {Vincent}, \citenamefont {Bœhm},\ and\
  \citenamefont {McCabe}}]{Wilkinson:2016gsy}%
  \BibitemOpen
  \bibfield  {author} {\bibinfo {author} {\bibfnamefont {R.~J.}\ \bibnamefont
  {Wilkinson}}, \bibinfo {author} {\bibfnamefont {A.~C.}\ \bibnamefont
  {Vincent}}, \bibinfo {author} {\bibfnamefont {C.}~\bibnamefont {Bœhm}}, \
  and\ \bibinfo {author} {\bibfnamefont {C.}~\bibnamefont {McCabe}},\ }\href
  {\doibase 10.1103/PhysRevD.94.103525} {\bibfield  {journal} {\bibinfo
  {journal} {Phys. Rev.}\ }\textbf {\bibinfo {volume} {D94}},\ \bibinfo {pages}
  {103525} (\bibinfo {year} {2016})},\ \Eprint
  {http://arxiv.org/abs/1602.01114} {arXiv:1602.01114 [astro-ph.CO]}
  \BibitemShut {NoStop}%
\bibitem [{\citenamefont {Beacom}\ and\ \citenamefont
  {Yuksel}(2006)}]{Beacom:2005qv}%
  \BibitemOpen
  \bibfield  {author} {\bibinfo {author} {\bibfnamefont {J.~F.}\ \bibnamefont
  {Beacom}}\ and\ \bibinfo {author} {\bibfnamefont {H.}~\bibnamefont
  {Yuksel}},\ }\href {\doibase 10.1103/PhysRevLett.97.071102} {\bibfield
  {journal} {\bibinfo  {journal} {Phys. Rev. Lett.}\ }\textbf {\bibinfo
  {volume} {97}},\ \bibinfo {pages} {071102} (\bibinfo {year} {2006})},\
  \Eprint {http://arxiv.org/abs/astro-ph/0512411} {arXiv:astro-ph/0512411
  [astro-ph]} \BibitemShut {NoStop}%
\bibitem [{\citenamefont {Radice}\ and\ \citenamefont
  {Rezzolla}(2012)}]{radice:2012cu}%
  \BibitemOpen
  \bibfield  {author} {\bibinfo {author} {\bibfnamefont {D.}~\bibnamefont
  {Radice}}\ and\ \bibinfo {author} {\bibfnamefont {L.}~\bibnamefont
  {Rezzolla}},\ }\href {\doibase 10.1051/0004-6361/201219735} {\bibfield
  {journal} {\bibinfo  {journal} {Astron. Astrophys.}\ }\textbf {\bibinfo
  {volume} {547}},\ \bibinfo {pages} {A26} (\bibinfo {year} {2012})},\ \Eprint
  {http://arxiv.org/abs/1206.6502} {arXiv:1206.6502 [astro-ph.IM]} \BibitemShut
  {NoStop}%
\bibitem [{\citenamefont {Radice}\ \emph
  {et~al.}(2014{\natexlab{a}})\citenamefont {Radice}, \citenamefont
  {Rezzolla},\ and\ \citenamefont {Galeazzi}}]{radice:2013hxh}%
  \BibitemOpen
  \bibfield  {author} {\bibinfo {author} {\bibfnamefont {D.}~\bibnamefont
  {Radice}}, \bibinfo {author} {\bibfnamefont {L.}~\bibnamefont {Rezzolla}}, \
  and\ \bibinfo {author} {\bibfnamefont {F.}~\bibnamefont {Galeazzi}},\ }\href
  {\doibase 10.1093/mnrasl/slt137} {\bibfield  {journal} {\bibinfo  {journal}
  {Mon. Not. Roy. Astron. Soc.}\ }\textbf {\bibinfo {volume} {437}},\ \bibinfo
  {pages} {L46} (\bibinfo {year} {2014}{\natexlab{a}})},\ \Eprint
  {http://arxiv.org/abs/1306.6052} {arXiv:1306.6052 [gr-qc]} \BibitemShut
  {NoStop}%
\bibitem [{\citenamefont {Radice}\ \emph
  {et~al.}(2014{\natexlab{b}})\citenamefont {Radice}, \citenamefont
  {Rezzolla},\ and\ \citenamefont {Galeazzi}}]{radice:2013xpa}%
  \BibitemOpen
  \bibfield  {author} {\bibinfo {author} {\bibfnamefont {D.}~\bibnamefont
  {Radice}}, \bibinfo {author} {\bibfnamefont {L.}~\bibnamefont {Rezzolla}}, \
  and\ \bibinfo {author} {\bibfnamefont {F.}~\bibnamefont {Galeazzi}},\ }\href
  {\doibase 10.1088/0264-9381/31/7/075012} {\bibfield  {journal} {\bibinfo
  {journal} {Class. Quant. Grav.}\ }\textbf {\bibinfo {volume} {31}},\ \bibinfo
  {pages} {075012} (\bibinfo {year} {2014}{\natexlab{b}})},\ \Eprint
  {http://arxiv.org/abs/1312.5004} {arXiv:1312.5004 [gr-qc]} \BibitemShut
  {NoStop}%
\bibitem [{\citenamefont {Steiner}\ \emph {et~al.}(2013)\citenamefont
  {Steiner}, \citenamefont {Hempel},\ and\ \citenamefont
  {Fischer}}]{steiner:2012rk}%
  \BibitemOpen
  \bibfield  {author} {\bibinfo {author} {\bibfnamefont {A.~W.}\ \bibnamefont
  {Steiner}}, \bibinfo {author} {\bibfnamefont {M.}~\bibnamefont {Hempel}}, \
  and\ \bibinfo {author} {\bibfnamefont {T.}~\bibnamefont {Fischer}},\ }\href
  {\doibase 10.1088/0004-637X/774/1/17} {\bibfield  {journal} {\bibinfo
  {journal} {Astrophys. J.}\ }\textbf {\bibinfo {volume} {774}},\ \bibinfo
  {pages} {17} (\bibinfo {year} {2013})},\ \Eprint
  {http://arxiv.org/abs/1207.2184} {arXiv:1207.2184 [astro-ph.SR]} \BibitemShut
  {NoStop}%
\bibitem [{\citenamefont {Radice}\ \emph {et~al.}(2016)\citenamefont {Radice},
  \citenamefont {Galeazzi}, \citenamefont {Lippuner}, \citenamefont {Roberts},
  \citenamefont {Ott},\ and\ \citenamefont {Rezzolla}}]{radice:2016dwd}%
  \BibitemOpen
  \bibfield  {author} {\bibinfo {author} {\bibfnamefont {D.}~\bibnamefont
  {Radice}}, \bibinfo {author} {\bibfnamefont {F.}~\bibnamefont {Galeazzi}},
  \bibinfo {author} {\bibfnamefont {J.}~\bibnamefont {Lippuner}}, \bibinfo
  {author} {\bibfnamefont {L.~F.}\ \bibnamefont {Roberts}}, \bibinfo {author}
  {\bibfnamefont {C.~D.}\ \bibnamefont {Ott}}, \ and\ \bibinfo {author}
  {\bibfnamefont {L.}~\bibnamefont {Rezzolla}},\ }\href {\doibase
  10.1093/mnras/stw1227} {\bibfield  {journal} {\bibinfo  {journal} {Mon. Not.
  Roy. Astron. Soc.}\ }\textbf {\bibinfo {volume} {460}},\ \bibinfo {pages}
  {3255} (\bibinfo {year} {2016})},\ \Eprint {http://arxiv.org/abs/1601.02426}
  {arXiv:1601.02426 [astro-ph.HE]} \BibitemShut {NoStop}%
\bibitem [{\citenamefont {Radice}\ \emph {et~al.}(2018)\citenamefont {Radice},
  \citenamefont {Perego}, \citenamefont {Hotokezaka}, \citenamefont {Fromm},
  \citenamefont {Bernuzzi},\ and\ \citenamefont {Roberts}}]{Radice:2018pdn}%
  \BibitemOpen
  \bibfield  {author} {\bibinfo {author} {\bibfnamefont {D.}~\bibnamefont
  {Radice}}, \bibinfo {author} {\bibfnamefont {A.}~\bibnamefont {Perego}},
  \bibinfo {author} {\bibfnamefont {K.}~\bibnamefont {Hotokezaka}}, \bibinfo
  {author} {\bibfnamefont {S.~A.}\ \bibnamefont {Fromm}}, \bibinfo {author}
  {\bibfnamefont {S.}~\bibnamefont {Bernuzzi}}, \ and\ \bibinfo {author}
  {\bibfnamefont {L.~F.}\ \bibnamefont {Roberts}},\ }\href@noop {} {\
  (\bibinfo {year} {2018})},\ \Eprint {http://arxiv.org/abs/1809.11161}
  {arXiv:1809.11161 [astro-ph.HE]} \BibitemShut {NoStop}%
\bibitem [{\citenamefont {Perego}\ \emph {et~al.}(2017)\citenamefont {Perego},
  \citenamefont {Radice},\ and\ \citenamefont {Bernuzzi}}]{Perego:2017wtu}%
  \BibitemOpen
  \bibfield  {author} {\bibinfo {author} {\bibfnamefont {A.}~\bibnamefont
  {Perego}}, \bibinfo {author} {\bibfnamefont {D.}~\bibnamefont {Radice}}, \
  and\ \bibinfo {author} {\bibfnamefont {S.}~\bibnamefont {Bernuzzi}},\ }\href
  {\doibase 10.3847/2041-8213/aa9ab9} {\bibfield  {journal} {\bibinfo
  {journal} {Astrophys. J.}\ }\textbf {\bibinfo {volume} {850}},\ \bibinfo
  {pages} {L37} (\bibinfo {year} {2017})},\ \Eprint
  {http://arxiv.org/abs/1711.03982} {arXiv:1711.03982 [astro-ph.HE]}
  \BibitemShut {NoStop}%
\bibitem [{\citenamefont {Rosswog}\ \emph {et~al.}(2014)\citenamefont
  {Rosswog}, \citenamefont {Korobkin}, \citenamefont {Arcones}, \citenamefont
  {Thielemann},\ and\ \citenamefont {Piran}}]{Rosswog:2013kqa}%
  \BibitemOpen
  \bibfield  {author} {\bibinfo {author} {\bibfnamefont {S.}~\bibnamefont
  {Rosswog}}, \bibinfo {author} {\bibfnamefont {O.}~\bibnamefont {Korobkin}},
  \bibinfo {author} {\bibfnamefont {A.}~\bibnamefont {Arcones}}, \bibinfo
  {author} {\bibfnamefont {F.~K.}\ \bibnamefont {Thielemann}}, \ and\ \bibinfo
  {author} {\bibfnamefont {T.}~\bibnamefont {Piran}},\ }\href {\doibase
  10.1093/mnras/stt2502} {\bibfield  {journal} {\bibinfo  {journal} {Mon. Not.
  Roy. Astron. Soc.}\ }\textbf {\bibinfo {volume} {439}},\ \bibinfo {pages}
  {744} (\bibinfo {year} {2014})},\ \Eprint {http://arxiv.org/abs/1307.2939}
  {arXiv:1307.2939 [astro-ph.HE]} \BibitemShut {NoStop}%
\bibitem [{\citenamefont {Kawaguchi}\ \emph {et~al.}(2016)\citenamefont
  {Kawaguchi}, \citenamefont {Kyutoku}, \citenamefont {Shibata},\ and\
  \citenamefont {Tanaka}}]{Kawaguchi:2016ana}%
  \BibitemOpen
  \bibfield  {author} {\bibinfo {author} {\bibfnamefont {K.}~\bibnamefont
  {Kawaguchi}}, \bibinfo {author} {\bibfnamefont {K.}~\bibnamefont {Kyutoku}},
  \bibinfo {author} {\bibfnamefont {M.}~\bibnamefont {Shibata}}, \ and\
  \bibinfo {author} {\bibfnamefont {M.}~\bibnamefont {Tanaka}},\ }\href
  {\doibase 10.3847/0004-637X/825/1/52} {\bibfield  {journal} {\bibinfo
  {journal} {Astrophys. J.}\ }\textbf {\bibinfo {volume} {825}},\ \bibinfo
  {pages} {52} (\bibinfo {year} {2016})},\ \Eprint
  {http://arxiv.org/abs/1601.07711} {arXiv:1601.07711 [astro-ph.HE]}
  \BibitemShut {NoStop}%
\bibitem [{\citenamefont {Barnes}\ \emph {et~al.}(2016)\citenamefont {Barnes},
  \citenamefont {Kasen}, \citenamefont {Wu},\ and\ \citenamefont
  {Martínez-Pinedo}}]{Barnes:2016umi}%
  \BibitemOpen
  \bibfield  {author} {\bibinfo {author} {\bibfnamefont {J.}~\bibnamefont
  {Barnes}}, \bibinfo {author} {\bibfnamefont {D.}~\bibnamefont {Kasen}},
  \bibinfo {author} {\bibfnamefont {M.-R.}\ \bibnamefont {Wu}}, \ and\ \bibinfo
  {author} {\bibfnamefont {G.}~\bibnamefont {Martínez-Pinedo}},\ }\href
  {\doibase 10.3847/0004-637X/829/2/110} {\bibfield  {journal} {\bibinfo
  {journal} {Astrophys. J.}\ }\textbf {\bibinfo {volume} {829}},\ \bibinfo
  {pages} {110} (\bibinfo {year} {2016})},\ \Eprint
  {http://arxiv.org/abs/1605.07218} {arXiv:1605.07218 [astro-ph.HE]}
  \BibitemShut {NoStop}%
\bibitem [{\citenamefont {Waxman}\ \emph {et~al.}(2017)\citenamefont {Waxman},
  \citenamefont {Ofek}, \citenamefont {Kushnir},\ and\ \citenamefont
  {Gal-Yam}}]{Waxman:2017sqv}%
  \BibitemOpen
  \bibfield  {author} {\bibinfo {author} {\bibfnamefont {E.}~\bibnamefont
  {Waxman}}, \bibinfo {author} {\bibfnamefont {E.~O.}\ \bibnamefont {Ofek}},
  \bibinfo {author} {\bibfnamefont {D.}~\bibnamefont {Kushnir}}, \ and\
  \bibinfo {author} {\bibfnamefont {A.}~\bibnamefont {Gal-Yam}},\ }\href@noop
  {} {\  (\bibinfo {year} {2017})},\ \Eprint {http://arxiv.org/abs/1711.09638}
  {arXiv:1711.09638 [astro-ph.HE]} \BibitemShut {NoStop}%
\bibitem [{\citenamefont {Qian}\ \emph {et~al.}(1993)\citenamefont {Qian},
  \citenamefont {Fuller}, \citenamefont {Mathews}, \citenamefont {Mayle},
  \citenamefont {Wilson},\ and\ \citenamefont {Woosley}}]{Qian:1993dg}%
  \BibitemOpen
  \bibfield  {author} {\bibinfo {author} {\bibfnamefont {Y.-Z.}\ \bibnamefont
  {Qian}}, \bibinfo {author} {\bibfnamefont {G.~M.}\ \bibnamefont {Fuller}},
  \bibinfo {author} {\bibfnamefont {G.~J.}\ \bibnamefont {Mathews}}, \bibinfo
  {author} {\bibfnamefont {R.}~\bibnamefont {Mayle}}, \bibinfo {author}
  {\bibfnamefont {J.~R.}\ \bibnamefont {Wilson}}, \ and\ \bibinfo {author}
  {\bibfnamefont {S.~E.}\ \bibnamefont {Woosley}},\ }\href {\doibase
  10.1103/PhysRevLett.71.1965} {\bibfield  {journal} {\bibinfo  {journal}
  {Phys. Rev. Lett.}\ }\textbf {\bibinfo {volume} {71}},\ \bibinfo {pages}
  {1965} (\bibinfo {year} {1993})}\BibitemShut {NoStop}%
\bibitem [{\citenamefont {Munz}()}]{Munz:1994}%
  \BibitemOpen
  \bibfield  {author} {\bibinfo {author} {\bibfnamefont {C.-D.}\ \bibnamefont
  {Munz}},\ }\href {\doibase 10.1002/mma.1670170803} {\bibfield  {journal}
  {\bibinfo  {journal} {Mathematical Methods in the Applied Sciences}\ }\textbf
  {\bibinfo {volume} {17}},\ \bibinfo {pages} {597}}\BibitemShut {NoStop}%
\bibitem [{\citenamefont {Duan}\ \emph {et~al.}(2006)\citenamefont {Duan},
  \citenamefont {Fuller}, \citenamefont {Carlson},\ and\ \citenamefont
  {Qian}}]{Duan:2006an}%
  \BibitemOpen
  \bibfield  {author} {\bibinfo {author} {\bibfnamefont {H.}~\bibnamefont
  {Duan}}, \bibinfo {author} {\bibfnamefont {G.~M.}\ \bibnamefont {Fuller}},
  \bibinfo {author} {\bibfnamefont {J.}~\bibnamefont {Carlson}}, \ and\
  \bibinfo {author} {\bibfnamefont {Y.-Z.}\ \bibnamefont {Qian}},\ }\href
  {\doibase 10.1103/PhysRevD.74.105014} {\bibfield  {journal} {\bibinfo
  {journal} {Phys. Rev.}\ }\textbf {\bibinfo {volume} {D74}},\ \bibinfo {pages}
  {105014} (\bibinfo {year} {2006})},\ \Eprint
  {http://arxiv.org/abs/astro-ph/0606616} {arXiv:astro-ph/0606616 [astro-ph]}
  \BibitemShut {NoStop}%
\bibitem [{sup()}]{suppmat}%
  \BibitemOpen
  \href@noop {} {\bibinfo  {journal} {See Supplemental Material [url], which
  includes Ref.~\cite{Waxman:2017sqv, Metzger:2016pju}, for further details
  regarding ejecta cooling as well as schematic representation of the
  atmospheric layer}\ }\BibitemShut {NoStop}%
\bibitem [{\citenamefont {Abbott}\ \emph {et~al.}(2016)\citenamefont {Abbott}
  \emph {et~al.}}]{Abbott:2016ymx}%
  \BibitemOpen
\bibfield  {journal} {  }\bibfield  {author} {\bibinfo {author} {\bibfnamefont
  {B.~P.}\ \bibnamefont {Abbott}} \emph {et~al.} (\bibinfo {collaboration}
  {Virgo, LIGO Scientific}),\ }\href {\doibase 10.3847/2041-8205/832/2/L21}
  {\bibfield  {journal} {\bibinfo  {journal} {Astrophys. J.}\ }\textbf
  {\bibinfo {volume} {832}},\ \bibinfo {pages} {L21} (\bibinfo {year}
  {2016})},\ \Eprint {http://arxiv.org/abs/1607.07456} {arXiv:1607.07456
  [astro-ph.HE]} \BibitemShut {NoStop}%
\bibitem [{\citenamefont {Mapelli}\ and\ \citenamefont
  {Giacobbo}(2018)}]{Mapelli:2018wys}%
  \BibitemOpen
  \bibfield  {author} {\bibinfo {author} {\bibfnamefont {M.}~\bibnamefont
  {Mapelli}}\ and\ \bibinfo {author} {\bibfnamefont {N.}~\bibnamefont
  {Giacobbo}},\ }\href {\doibase 10.1093/mnras/sty1613} {\  (\bibinfo {year}
  {2018}),\ 10.1093/mnras/sty1613},\ \Eprint {http://arxiv.org/abs/1806.04866}
  {arXiv:1806.04866 [astro-ph.HE]} \BibitemShut {NoStop}%
\bibitem [{\citenamefont {Chruslinska}\ \emph {et~al.}(2018)\citenamefont
  {Chruslinska}, \citenamefont {Belczynski}, \citenamefont {Klencki},\ and\
  \citenamefont {Benacquista}}]{Chruslinska:2017odi}%
  \BibitemOpen
  \bibfield  {author} {\bibinfo {author} {\bibfnamefont {M.}~\bibnamefont
  {Chruslinska}}, \bibinfo {author} {\bibfnamefont {K.}~\bibnamefont
  {Belczynski}}, \bibinfo {author} {\bibfnamefont {J.}~\bibnamefont {Klencki}},
  \ and\ \bibinfo {author} {\bibfnamefont {M.}~\bibnamefont {Benacquista}},\
  }\href {\doibase 10.1093/mnras/stx2923} {\bibfield  {journal} {\bibinfo
  {journal} {Mon. Not. Roy. Astron. Soc.}\ }\textbf {\bibinfo {volume} {474}},\
  \bibinfo {pages} {2937} (\bibinfo {year} {2018})},\ \Eprint
  {http://arxiv.org/abs/1708.07885} {arXiv:1708.07885 [astro-ph.HE]}
  \BibitemShut {NoStop}%
\bibitem [{\citenamefont {Jean}\ \emph {et~al.}(2009)\citenamefont {Jean},
  \citenamefont {Gillard}, \citenamefont {Marcowith},\ and\ \citenamefont
  {Ferriere}}]{Jean:2009zj}%
  \BibitemOpen
  \bibfield  {author} {\bibinfo {author} {\bibfnamefont {P.}~\bibnamefont
  {Jean}}, \bibinfo {author} {\bibfnamefont {W.}~\bibnamefont {Gillard}},
  \bibinfo {author} {\bibfnamefont {A.}~\bibnamefont {Marcowith}}, \ and\
  \bibinfo {author} {\bibfnamefont {K.}~\bibnamefont {Ferriere}},\ }\href
  {\doibase 10.1051/0004-6361/200809830} {\bibfield  {journal} {\bibinfo
  {journal} {Astron. Astrophys.}\ }\textbf {\bibinfo {volume} {508}},\ \bibinfo
  {pages} {1099} (\bibinfo {year} {2009})},\ \Eprint
  {http://arxiv.org/abs/0909.4022} {arXiv:0909.4022 [astro-ph.HE]} \BibitemShut
  {NoStop}%
\bibitem [{\citenamefont {Berger}(2014)}]{Berger:2013jza}%
  \BibitemOpen
  \bibfield  {author} {\bibinfo {author} {\bibfnamefont {E.}~\bibnamefont
  {Berger}},\ }\href {\doibase 10.1146/annurev-astro-081913-035926} {\bibfield
  {journal} {\bibinfo  {journal} {Ann. Rev. Astron. Astrophys.}\ }\textbf
  {\bibinfo {volume} {52}},\ \bibinfo {pages} {43} (\bibinfo {year} {2014})},\
  \Eprint {http://arxiv.org/abs/1311.2603} {arXiv:1311.2603 [astro-ph.HE]}
  \BibitemShut {NoStop}%
\bibitem [{\citenamefont {Tsang}(2013)}]{Tsang:2013mca}%
  \BibitemOpen
  \bibfield  {author} {\bibinfo {author} {\bibfnamefont {D.}~\bibnamefont
  {Tsang}},\ }\href {\doibase 10.1088/0004-637X/777/2/103} {\bibfield
  {journal} {\bibinfo  {journal} {Astrophys. J.}\ }\textbf {\bibinfo {volume}
  {777}},\ \bibinfo {pages} {103} (\bibinfo {year} {2013})},\ \Eprint
  {http://arxiv.org/abs/1307.3554} {arXiv:1307.3554 [astro-ph.HE]} \BibitemShut
  {NoStop}%
\bibitem [{\citenamefont {Churazov}\ \emph {et~al.}(2011)\citenamefont
  {Churazov}, \citenamefont {Sazonov}, \citenamefont {Tsygankov}, \citenamefont
  {Sunyaev},\ and\ \citenamefont {Varshalovich}}]{Churazov:2010wy}%
  \BibitemOpen
  \bibfield  {author} {\bibinfo {author} {\bibfnamefont {E.}~\bibnamefont
  {Churazov}}, \bibinfo {author} {\bibfnamefont {S.}~\bibnamefont {Sazonov}},
  \bibinfo {author} {\bibfnamefont {S.}~\bibnamefont {Tsygankov}}, \bibinfo
  {author} {\bibfnamefont {R.}~\bibnamefont {Sunyaev}}, \ and\ \bibinfo
  {author} {\bibfnamefont {D.}~\bibnamefont {Varshalovich}},\ }\href {\doibase
  10.1111/j.1365-2966.2010.17804.x} {\bibfield  {journal} {\bibinfo  {journal}
  {Mon. Not. Roy. Astron. Soc.}\ }\textbf {\bibinfo {volume} {411}},\ \bibinfo
  {pages} {1727} (\bibinfo {year} {2011})},\ \Eprint
  {http://arxiv.org/abs/1010.0864} {arXiv:1010.0864 [astro-ph.HE]} \BibitemShut
  {NoStop}%
\bibitem [{\citenamefont {Li}\ \emph {et~al.}(2011)\citenamefont {Li},
  \citenamefont {Chornock}, \citenamefont {Leaman}, \citenamefont {Filippenko},
  \citenamefont {Poznanski}, \citenamefont {Wang}, \citenamefont
  {Ganeshalingam},\ and\ \citenamefont {Mannucci}}]{Li:2010kd}%
  \BibitemOpen
  \bibfield  {author} {\bibinfo {author} {\bibfnamefont {W.}~\bibnamefont
  {Li}}, \bibinfo {author} {\bibfnamefont {R.}~\bibnamefont {Chornock}},
  \bibinfo {author} {\bibfnamefont {J.}~\bibnamefont {Leaman}}, \bibinfo
  {author} {\bibfnamefont {A.~V.}\ \bibnamefont {Filippenko}}, \bibinfo
  {author} {\bibfnamefont {D.}~\bibnamefont {Poznanski}}, \bibinfo {author}
  {\bibfnamefont {X.}~\bibnamefont {Wang}}, \bibinfo {author} {\bibfnamefont
  {M.}~\bibnamefont {Ganeshalingam}}, \ and\ \bibinfo {author} {\bibfnamefont
  {F.}~\bibnamefont {Mannucci}},\ }\href {\doibase
  10.1111/j.1365-2966.2011.18162.x} {\bibfield  {journal} {\bibinfo  {journal}
  {Mon. Not. Roy. Astron. Soc.}\ }\textbf {\bibinfo {volume} {412}},\ \bibinfo
  {pages} {1473} (\bibinfo {year} {2011})},\ \Eprint
  {http://arxiv.org/abs/1006.4613} {arXiv:1006.4613 [astro-ph.SR]} \BibitemShut
  {NoStop}%
\bibitem [{\citenamefont {Adams}\ \emph {et~al.}(2013)\citenamefont {Adams},
  \citenamefont {Kochanek}, \citenamefont {Beacom}, \citenamefont {Vagins},\
  and\ \citenamefont {Stanek}}]{Adams:2013ana}%
  \BibitemOpen
  \bibfield  {author} {\bibinfo {author} {\bibfnamefont {S.~M.}\ \bibnamefont
  {Adams}}, \bibinfo {author} {\bibfnamefont {C.~S.}\ \bibnamefont {Kochanek}},
  \bibinfo {author} {\bibfnamefont {J.~F.}\ \bibnamefont {Beacom}}, \bibinfo
  {author} {\bibfnamefont {M.~R.}\ \bibnamefont {Vagins}}, \ and\ \bibinfo
  {author} {\bibfnamefont {K.~Z.}\ \bibnamefont {Stanek}},\ }\href {\doibase
  10.1088/0004-637X/778/2/164} {\bibfield  {journal} {\bibinfo  {journal}
  {Astrophys. J.}\ }\textbf {\bibinfo {volume} {778}},\ \bibinfo {pages} {164}
  (\bibinfo {year} {2013})},\ \Eprint {http://arxiv.org/abs/1306.0559}
  {arXiv:1306.0559 [astro-ph.HE]} \BibitemShut {NoStop}%
\bibitem [{\citenamefont {Abbott}\ \emph
  {et~al.}(2017{\natexlab{b}})\citenamefont {Abbott} \emph
  {et~al.}}]{TheLIGOScientific:2017qsa}%
  \BibitemOpen
  \bibfield  {author} {\bibinfo {author} {\bibfnamefont {B.}~\bibnamefont
  {Abbott}} \emph {et~al.} (\bibinfo {collaboration} {Virgo, LIGO
  Scientific}),\ }\href {\doibase 10.1103/PhysRevLett.119.161101} {\bibfield
  {journal} {\bibinfo  {journal} {Phys. Rev. Lett.}\ }\textbf {\bibinfo
  {volume} {119}},\ \bibinfo {pages} {161101} (\bibinfo {year}
  {2017}{\natexlab{b}})},\ \Eprint {http://arxiv.org/abs/1710.05832}
  {arXiv:1710.05832 [gr-qc]} \BibitemShut {NoStop}%
\bibitem [{\citenamefont {Ji}\ \emph {et~al.}(2016)\citenamefont {Ji},
  \citenamefont {Frebel}, \citenamefont {Chiti},\ and\ \citenamefont
  {Simon}}]{Ji:2015wzg}%
  \BibitemOpen
  \bibfield  {author} {\bibinfo {author} {\bibfnamefont {A.~P.}\ \bibnamefont
  {Ji}}, \bibinfo {author} {\bibfnamefont {A.}~\bibnamefont {Frebel}}, \bibinfo
  {author} {\bibfnamefont {A.}~\bibnamefont {Chiti}}, \ and\ \bibinfo {author}
  {\bibfnamefont {J.~D.}\ \bibnamefont {Simon}},\ }\href {\doibase
  10.1038/nature17425} {\bibfield  {journal} {\bibinfo  {journal} {Nature}\
  }\textbf {\bibinfo {volume} {531}},\ \bibinfo {pages} {610} (\bibinfo {year}
  {2016})},\ \Eprint {http://arxiv.org/abs/1512.01558} {arXiv:1512.01558
  [astro-ph.GA]} \BibitemShut {NoStop}%
\bibitem [{\citenamefont {Beniamini}\ \emph {et~al.}(2018)\citenamefont
  {Beniamini}, \citenamefont {Dvorkin},\ and\ \citenamefont
  {Silk}}]{Beniamini:2017qqy}%
  \BibitemOpen
  \bibfield  {author} {\bibinfo {author} {\bibfnamefont {P.}~\bibnamefont
  {Beniamini}}, \bibinfo {author} {\bibfnamefont {I.}~\bibnamefont {Dvorkin}},
  \ and\ \bibinfo {author} {\bibfnamefont {J.}~\bibnamefont {Silk}},\ }\href
  {\doibase 10.1093/mnras/sty1035} {\bibfield  {journal} {\bibinfo  {journal}
  {Mon. Not. Roy. Astron. Soc.}\ }\textbf {\bibinfo {volume} {478}},\ \bibinfo
  {pages} {1994} (\bibinfo {year} {2018})},\ \Eprint
  {http://arxiv.org/abs/1711.02683} {arXiv:1711.02683 [astro-ph.HE]}
  \BibitemShut {NoStop}%
\end{thebibliography}%

\vspace{10em}

\newpage
\begin{titlepage}
\begin{center}
{\large{\bf Supplemental Material}}
\end{center}
\vspace{3em}
\end{titlepage}

\section{Ejecta Cooling}

A better understanding of the ejecta cooling can be obtained from analytic kilonova models, fit to light-curve data \cite{Metzger:2016pju,Waxman:2017sqv}.~In spherically expanding ejecta, the distribution of mass with velocity greater than $v$ is effectively described as $M_v = M (v/v_0)^{-\beta}$, with $v \geq v_0$. Here, $v_0 \sim 0.1$ is the average ($\approx$ minimum) velocity, $M$ is the total ejecta mass and $\beta \sim \mathcal{O}(1)$ is a phenomenological parameter.

The thermal energy in the ejecta layer expanding with velocity $v$ and mass $dM_v$ evolves as  
\begin{equation} \label{eq:dedt}
\dfrac{d E_v}{dt} = - \dfrac{E_v}{R_v}\dfrac{d R_v}{dt} - L_v + \dot{Q}~,
\end{equation}
where the first term on the right hand side accounts for adiabatic losses and $R_v = v t$ denotes the layer radius. Here, the
\begin{equation} 
L_v = \dfrac{E_v}{t_{d,v} + t_{lc,v}} 
\end{equation}
term accounts for radiative losses, with $t_{d,v}$ describing the diffusion time
\begin{equation}
t_{d,v} \simeq \dfrac{M_v^{1/4} \kappa_v}{4 \pi M v_0 t}~,
\end{equation}
where $\kappa_v$ is the photon opacity and $v_0$ is the initial velocity. The light-crossing timescale $t_{lc,v} = R_v = v t$ limits $L_v$, thus avoiding contributions from super-luminous photon propagation. This is particularly important when the ejecta layer is in the optically thin regime, i.e. when $t_{d,v} \ll 1$.  The $\dot{Q}$ term accounts for multiple ejecta heating sources, including $r$-process nucleosynthesis, central engine, fall-back accretion and a possible additional thin neutron layer.

Focusing solely on the adiabatic and radiative cooling in the outer layer, we neglect the $\dot{Q}$ heating terms. Since we are interested in the optically thin regime where $t_{d,v} \ll 1$, this term can also be neglected.
Hence, Eq.~\eqref{eq:dedt} simplifies to 
\begin{equation} \label{eq:ecool}
\dfrac{d E_v}{dt} = - \dfrac{E_v}{t} - \dfrac{E_v}{v t}~.
\end{equation}
Since $v \sim \mathcal{O}(0.3)$, integration yields
\begin{equation}
E_2 \simeq \dfrac{t_1}{t_2} E_1~.
\end{equation}
Translating thermal energy to temperature via
\begin{equation}
T_v \simeq \Big(\dfrac{3 E_v}{4 \pi a R_v^3}\Big)^{1/4}~,
\end{equation}
where $a$ is a parameter, we obtain
\begin{equation} \label{eq:tcool}
T_2 \simeq \dfrac{t_1}{t_2} T_1~.
\end{equation}
Thus, cooling of the ejecta is linear in time and without additional heating contributions the original temperature in the thin outer layers of the expanding ejecta will decrease quickly.

\section{Atmospheric Layer}

\begin{figure}[h]
      \includegraphics[width=0.45\textwidth]{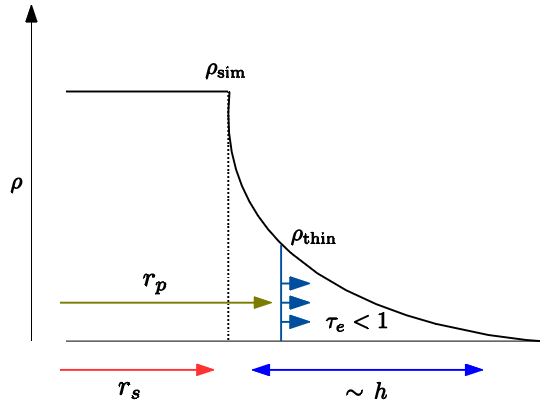}
\caption{Schematic representation of the ejecta outer layer profile soon after the merger. An atmospheric layer of characteristic thickness $\sim h$ and exponentially decreasing density lies beyond the surface of density $\rho_{\rm sim}$ as resolved by initial merger simulations. For atmospheric layers with density below $\rho_{\rm thin}$, which sets the boundary of the positron-sphere $r_p$, the ejecta is optically thin to positrons (i.e. $\tau_e < 1$). }
\label{fig:atmprofile} 
\end{figure} 

\begin{figure}[h]
      \includegraphics[width=0.35\textwidth]{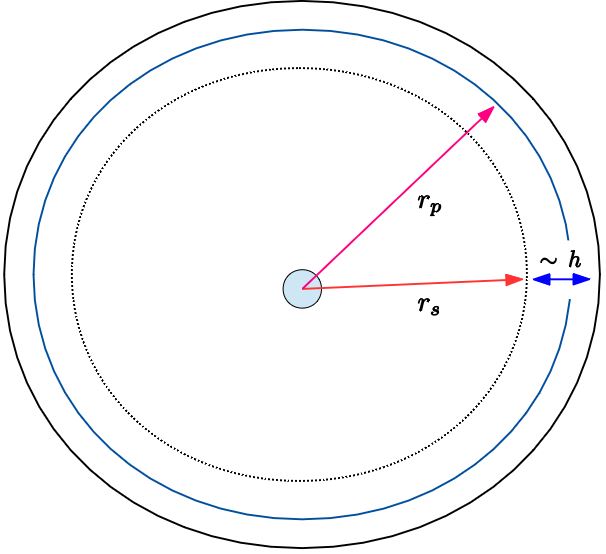}
\caption{Schematic representation of the ejecta soon after the merger. Merger remnant (grey, center), ejecta radius $r_s$ as resolved within initial merger simulations, positron-sphere $r_p$ as well as the outer atmospheric layer of characteristic size $h$ are displayed.}
\label{fig:ejecta_layers} 
\end{figure}

\vspace{20em}

\end{document}